\documentclass[preprint,3p,times,twocolumn,authoryear]{elsarticle}
\usepackage{amssymb}
\usepackage{amsmath}
\journal{Advances in Space Research}
\usepackage{lineno}
\modulolinenumbers[1]
\usepackage{xcolor}
\usepackage[citebordercolor=white]{hyperref}
\hypersetup{colorlinks=true,linkcolor={blue},citecolor={blue},urlcolor={red}}

\begin{document}
\begin{frontmatter}
\title{A generalized method for estimating solar wind speeds and densities using spectral broadening for a Kolmogorov turbulence spectrum} %% Article title

\author[label1]{Keshav Aggarwal\corref{cor1}}
\author[label2]{R. K. Choudhary}
\author[label1]{Abhirup Datta}
\author[label3]{Roopa M. V.}
\author[label4]{Takeshi Imamura}
\author[label5]{Hiroki Ando}

%% Author affiliations
\affiliation[label1]{organization={Department of Astronomy, Astrophysics and Space Engineering (DAASE), Indian Institute of Technology Indore}, addressline={Khandwa Road, Simrol}, city={Indore}, postcode={453552}, state={Madhya Pradesh}, country={India}}

\affiliation[label2]{organization={Space Physics Laboratory (SPL), Indian Space Research Organization}, addressline={Vikram Sarabhai Space Centre}, city={Thiruvananthapuram}, postcode={695022}, state={Kerala}, country={India}}

\affiliation[label3]{organization={ISRO Telemetry Tracking and Command Network (ISTRAC)}, city={Bengaluru}, postcode={560058}, state={Karnataka}, country={India}}

\affiliation[label4]{organization={Graduate School of Frontier Sciences, The University of Tokyo}, addressline={Kiban-tou 4H7, 5-1-5 Kashiwanoha}, city={Kashiwa}, postcode={277-8561}, state={Chiba}, country={Japan}}

\affiliation[label5]{organization={Faculty of Science, Kyoto Sangyo University},
addressline={}, city={Kyoto}, postcode={603-8555}, country={Japan}}

\cortext[cor1]{Corresponding author: keshavagg1098@gmail.com}

%% Abstract
\begin{abstract}
We present a unified method to derive both solar wind velocities and coronal electron densities in the near-Sun corona using Doppler spectral broadening of spacecraft radio signals. The method is generalized to be frequency independent under the assumption that electron density fluctuations follow a Kolmogorov spectrum. We validate the approach using S-band data from India's Mars Orbiter Mission during the October 2021 superior conjunction at 5-8 R$_\odot$, and X-band data from Japan's Akatsuki during June 2016 and October 2022 conjunctions spanning 1.4-10 R$_\odot$. From S-band we obtained wind speeds of 100-150 km s$^{-1}$ and electron densities of order $10^{10}$ m$^{-3}$. X-band results show speeds ranging from $\sim$150 km s$^{-1}$ near the equator to $\sim$400 km s$^{-1}$ in coronal-hole regions, with consistent radial trends in density. We provide a compact, frequency-scaled relation that maps Doppler spectral width to both $v$ and $N_e$. The formulation enables consistent application across telecommunication bands and complements in-situ probes for coronal plasma studies.
\end{abstract}

%%Research highlights
% \begin{highlights}
% \item Research highlight 1
% \item Research highlight 2
% \end{highlights}

%% Keywords

\end{frontmatter}

%% Add \usepackage{lineno} before \begin{document} and uncomment 
%% following line to enable line numbers
%% \linenumbers

%% main text
%%

%% Use \section commands to start a section
\section{Introduction}
The solar corona is one of the most dynamic plasma environments in the heliosphere, where the Sun's magnetic field and energetic particles transform the subsonic coronal plasma into the supersonic solar wind. This continuous outflow of charged particles, first theoretically predicted by \citet{Parker1958} and confirmed by early space missions \citep{Gringauz1960}, carries mass, momentum, and magnetic flux throughout interplanetary space, fundamentally shaping phenomena from planetary magnetospheric dynamics to cosmic ray modulation \citep{Tsurutani1992, Gopalswamy2001}. Despite decades of intensive study, the mechanisms governing coronal heating and solar wind acceleration remain incompletely understood, primarily due to the inherent difficulty of obtaining direct measurements in the extreme plasma environment near the Sun.

The coronal plasma environment is magnetically structured and dominated by free electrons that strongly influence radio wave propagation, making radio remote sensing a uniquely powerful diagnostic tool for probing near-Sun conditions. Radio occultation (RO) experiments exploit this fundamental plasma-electromagnetic coupling: when spacecraft pass behind the Sun from Earth's perspective during superior conjunction, their radio signals traverse the corona and undergo refraction, scattering, and Doppler shifting by the intervening plasma \citep{Muhleman1977, Tyler1977}. Analysis of the received signal's phase, amplitude, and frequency spectra enables inference of critical plasma properties, including electron density, turbulence characteristics, and bulk flow velocities \citep{Coles1989, Bird1982}.

Radio occultation has a distinguished heritage in solar physics, with pioneering studies utilizing signals from Mariner, Helios, and Voyager missions establishing fundamental relationships between observed signal characteristics and coronal plasma parameters \citep{Woo1978,Paetzold1987, Coles1991, Wohlmuth2001}. The technique gained prominence through systematic investigations of solar wind velocities via interplanetary scintillation \citep{Kojima1987,Tokumaru2012} and electron density measurements through precise time-delay observations \citep{Muhleman1981,Esposito1980}. These early efforts demonstrated RO's unique capability to extend plasma diagnostics closer to the Sun than achievable by in-situ probes, while providing coverage during geometric configurations inaccessible to dedicated solar missions.
Contemporary spacecraft continue to underscore RO's scientific value for coronal investigations. India's Mars Orbiter Mission (MOM) and Japan's Akatsuki Venus Climate Orbiter have furnished extensive long-baseline tracking datasets during superior conjunctions, enabling measurements at heliocentric distances of just a few solar radii-regions otherwise extremely challenging to study directly \citep{Imamura2017,Aggarwal2025a}. These modern observations have revealed the corona's complex, dynamic character, with solar wind speeds ranging from 100-150 km s$^{-1}$ in equatorial streamer regions to 400-600 km s$^{-1}$ in polar coronal holes \citep{McComas2008,Wang1990}, while electron densities exhibit characteristic radial gradients with significant modulations due to transient phenomena \citep{Leblanc1998, Mancuso2013}.

Previous RO analyses have developed empirical relations linking Doppler spectral broadening to solar wind properties, but these approaches typically remain constrained to specific frequency bands. Our earlier investigations utilized S-band signals from MOM's October 2021 conjunction to establish relationships between spectral width and solar wind velocity, yielding wind speeds of 100-150 km s$^{-1}$ and electron densities of approximately $10^{10}$ m$^{-3}$ at 5-8 R$_\odot$ \citep{Aggarwal2025a}. Complementary analysis of X-band signals from Akatsuki during June 2016 and October 2022 conjunctions captured wind speeds ranging from $\sim$150 km s$^{-1}$ near equatorial regions to $\sim$400 km s$^{-1}$ at higher latitudes associated with coronal holes \citep{Aggarwal2025b}. However, a persistent limitation of existing methodologies lies in their frequency-specific algorithm adjustments. Spacecraft employ diverse telecommunication bands-S-band (2-4 GHz), X-band (8-12 GHz), Ka-band, and beyond-each exhibiting distinct propagation characteristics and sensitivity to different aspects of coronal turbulence \citep{Armstrong1981, Morabito2009, Efimov2015}. This frequency dependence has constrained most studies to band-specific interpretations, complicating direct cross-mission comparisons and unified physical understanding.

To address these challenges, we present a generalized framework for simultaneously deriving solar wind velocities and coronal electron densities from Doppler spectral broadening observations. Our approach is grounded in the assumption that coronal electron density fluctuations follow a Kolmogorov turbulence spectrum, providing robust theoretical foundations for unifying observations across different radio frequencies \citep{Coles1989, Armstrong1981}. By using our relations derived from our S-band (MOM) and X-band (Akatsuki), we establish a compact relation that maps observed Doppler spectral width directly to radial solar wind velocity ($v$) and electron density ($N_e$) for any telecommunication frequency employed in spacecraft tracking in the radio regime. This unified diagnostic framework promises to streamline coronal plasma analysis across current and future missions, enabling consistent interpretation of diverse RO datasets and maximizing scientific return from routine spacecraft telecommunications. 

The structure of this paper proceeds as follows: Section 2 describes the MOM and Akatsuki datasets and superior conjunction observing geometries; our methodology and derivation of the generalized frequency relation; Section 3 presents comparative results for velocities and electron densities from both missions alongside empirical coronal model comparisons; and Section 5 summarizes conclusions while highlighting future research directions, including the need for more sophisticated treatment of turbulence spectrum effects.

\begin{figure*}[h]
 \includegraphics[width = 1\linewidth]{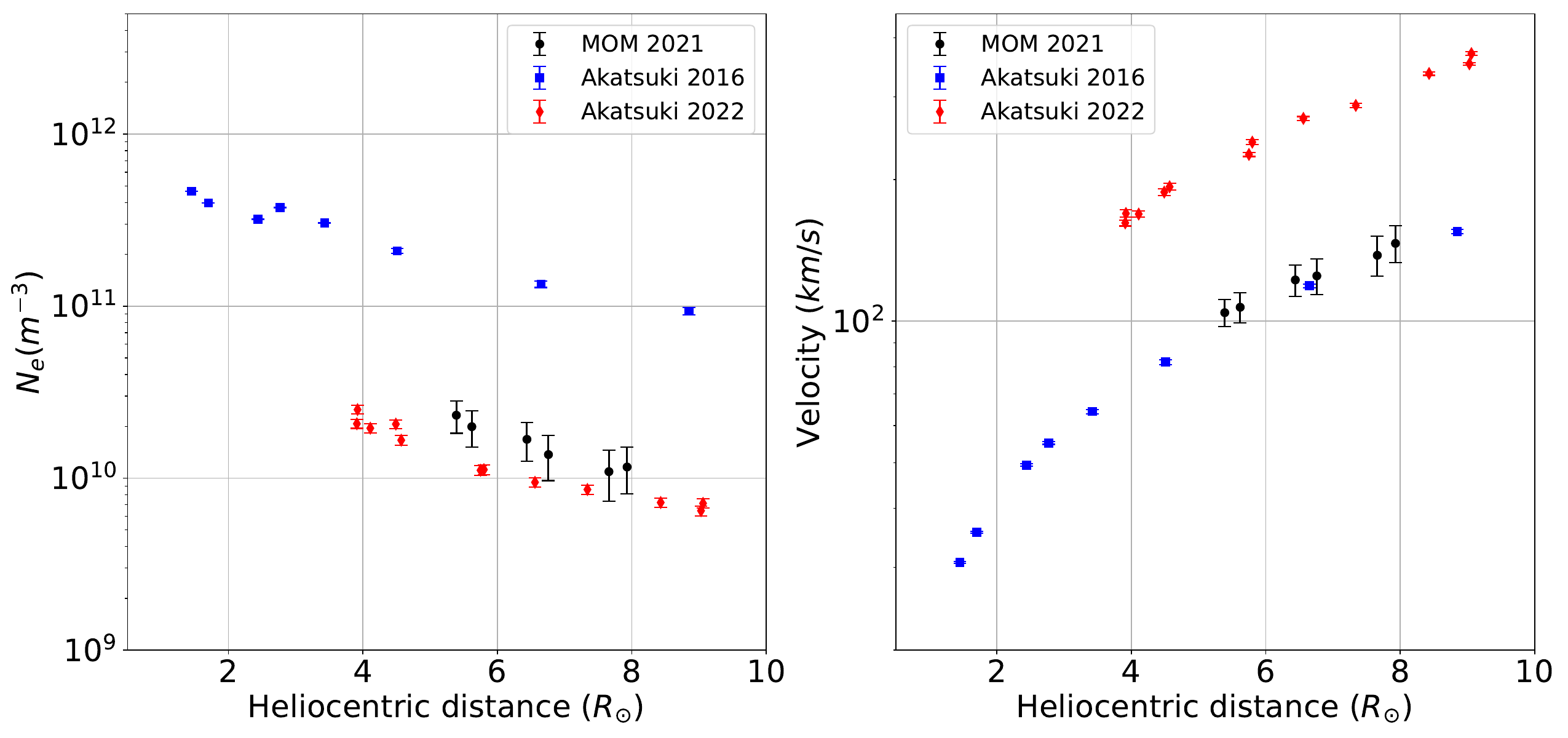}
 \caption{Radial profiles of the electron density and solar wind velocity derived from radio occultation experiments conducted with Akatsuki (2016, 2022) and the Mars Orbiter Mission (MOM, 2021). Left panel: Electron density variation ($N_e$) as a function of heliocentric distance, showing consistency across different epochs and spacecraft geometries. 
Right panel: Solar wind velocity profiles obtained simultaneously from the same set of occultations, demonstrating the comparative acceleration trends and variability with solar offset.}
\label{fig:ne-vperp}
\end{figure*}

\section{Method}

\subsection{MOM and Akatsuki Missions}

The Indian Mars Orbiter Mission (MOM), launched in 2013, successfully entered Martian orbit after a 300-day cruise and far exceeded its intended six-month mission lifetime, operating for more than eight years until September 2022 \citep{Arunan2015, Bhardwaj2016, Jain2022}. During the solar conjunction of October 2021, MOM was used to conduct radio occultation experiments while in an elliptical orbit around Mars. Telemetry, tracking, and command signals were transmitted via the spacecraft's 2.2 m high-gain parabolic offset antenna, which employed oscillators with an Allan variance of order $10^{-11}$ \citep{Ramamurthy2015}. The downlinked S-band (2.29 GHz) signals were recorded by the 32 m antenna at the Indian Deep Space Network (IDSN) in Bangalore using open-loop radio science receivers, similar to those described by \citet{Bedrossian2019}. S-band frequencies are well-suited for such studies because they lie within a transparent atmospheric window, ensuring minimal attenuation or scattering as signals propagate through Earth's atmosphere. As the signals traversed the solar corona during the October 2021 occultation event, they acquired characteristic signatures from plasma interactions, making them suitable for probing near-Sun conditions at heliocentric distances of 5–8 $R_\odot$.

Complementary observations were obtained with the Japanese Venus Orbiter Akatsuki, which has been engaged in solar occultation studies since its orbital insertion \citep{Chiba2022, Chiba2023, Jain_2023, Jain_2024}. For this work we considered data from two solar conjunctions: June 2016, during the descending phase of Solar Cycle 24 with low activity, and October 2022, during the ascending phase of Solar Cycle 25 with moderate to high activity. The spacecraft communicated with Earth using a 1.6 m high-gain radial line slot antenna, with oscillators exhibiting Allan variance of order $10^{-12}$ \citep{Oshima2011, Imamura2017}. X-band (8.41 GHz) signals were received at the 64 m antenna of the Usuda Deep Space Center in Japan \citep{Imamura2011} and by the Indian Deep Space Network in Byalalu, India \citep{Jain_2023}. Open-loop receiver systems recorded the signals in CCSDS-RDEF format, which were subsequently processed using standard pipelines involving carrier extraction, Fourier transformation, and fitting of Doppler power spectral densities to obtain measures of spectral broadening \citep{Tripathi2022, Aggarwal2025b}. 

Figure \ref{fig:ne-vperp} shows radial profiles of the electron density and solar wind velocity derived from radio occultation experiments conducted with Akatsuki (2016, 2022) (shown in blue and red respectively) and the Mars Orbiter Mission (MOM, 2021) (shown in black). Left panel shows electron density ($N_e$) as a function of heliocentric distance, while the right panel shows solar wind velocity profiles obtained simultaneously from the same set of occultations, demonstrating the comparative acceleration trends and variability with solar offset. As can be seen from the Fig \ref{fig:ne-vperp}, the high values of electron densities in 2016 correspond to lower velocities in the companion plot and the opposite situation is observed in the results from 2022, when the signals traversed the region near a coronal hole. However, it is also to be noted that for the 2021 experiment done using MOM that while the densities are not much higher as compared to the other measurements, the velocities are not that different from slow winds, leading us to assume that the region being probed was near streamer belts.

The unique orbital configurations of MOM and Akatsuki allowed radio line-of-sight geometries that probed the solar corona over a wide range of heliocentric distances, from 1.4 $R_\odot$ in the case of Akatsuki to 8 $R_\odot$ with MOM. As spacecraft radio signals propagated through the ionized plasma of the corona, they experienced amplitude scintillation, phase fluctuations, Doppler shifts, spectral broadening, and Faraday rotation. Analysis of these effects enables quantitative estimates of turbulence levels, electron density spectra, solar wind velocities, and magnetic field properties in the near-Sun corona \citep{Woo1976, Woo1977, Wexler2019, Wexler_2021}.

\subsection{Data sets and observing geometry}
Below we summarize, in a concise and self-contained manner, the observational and processing details for the two spacecraft data sets used to validate the method: India's Mars Orbiter Mission (MOM; S-band, October 2021) and Japan's Akatsuki (X-band, June 2016 and October 2022). Wherever precise station or instrument parameters are needed we cite earlier descriptions and note assumptions explicitly.
\begin{figure*}[ht]
\centering
\includegraphics[width = 1\linewidth]{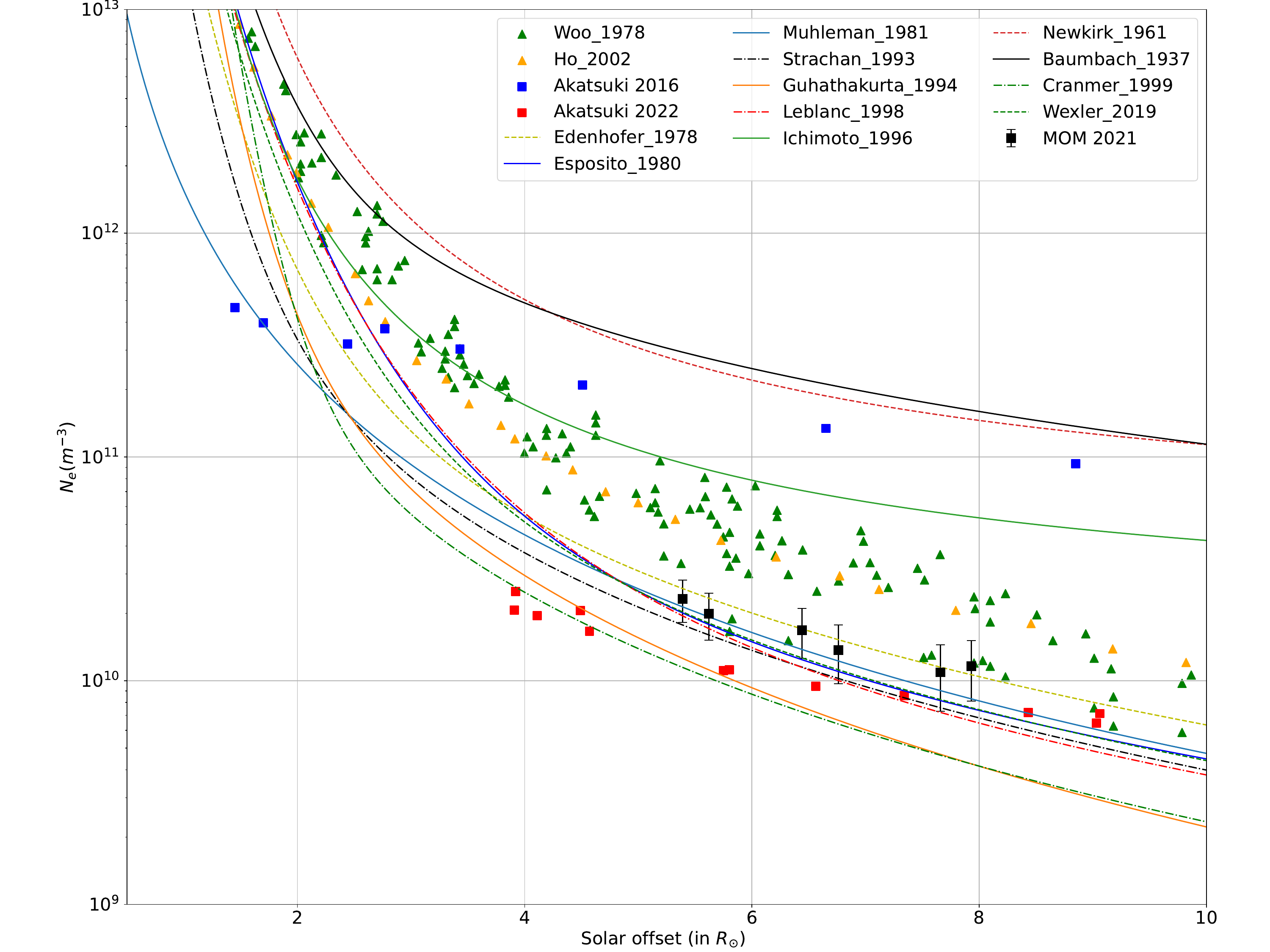}
\caption{Electron density estimates ($N_e$) derived using our unified Doppler spectral broadening technique for both the Mars Orbiter Mission (MOM) and Akatsuki radio occultation (RO) experiments are shown as a function of solar offset ($R_\odot$) compared against other studies in the literature.}
\label{fig:ne}
\end{figure*}

\noindent\textbf{Link frequencies:} MOM downlinks at S–band (nominally $\sim$2.29 GHz). Akatsuki uses X–band (nominally $\sim$8.41 GHz) \citep{Ramamurthy2015, Imamura2011, Oshima2011}.

\noindent\textbf{Heliocentric impact parameters and geometry:} MOM occultation rays sampled the near–Sun corona at heliocentric distances of approximately $5$-$8 R_{\odot}$ and were close to the ecliptic plane during the October 2021 superior conjunction \citep{Jain2022}. Akatsuki conjunction geometries probed a wider range, roughly $1.4$-$10 R_{\odot}$, sampling both equatorial streamer regions and coronal–hole latitudes in the 2016 (low activity) and 2022 (moderate/high activity) events \citep{Chiba2022, Jain_2024}.

\noindent\textbf{Timing and cadence:} Observations used 1 s raw frames carrier tracking which were averaged with a one–minute moving window to reduce short–term scatter following established practice \citep{Woo1979, Coles1991, Aggarwal2025a}.

\noindent\textbf{Station receivers and stability:} MOM S–band was received at the IDSN 32 m station (Bangalore) using open–loop radio science receivers; Akatsuki X–band was recorded at Usuda (64 m) and also by IDSN/Byalalu, archived in CCSDS–RDEF format \citep{Imamura2011, Jain_2024}. Oscillator Allan variances are of order $10^{-11}$ (MOM) and $10^{-12}$ (Akatsuki), and their small contribution to frequency error is explicitly accounted for in the uncertainty budget \citep{Ramamurthy2015, Oshima2011, Tripathi2022b}.

\noindent\textbf{Minimum impact parameters, latitudes and approximate times:} For MOM (Oct 2021) the minimum solar offset was $\sim5 R_{\odot}$ near ecliptic latitudes; for Akatsuki (Jun 2016 and Oct 2022) minimum offsets reached $\sim1.4 R_{\odot}$ sampling low–to–mid heliographic latitudes. Exact station–time stamps and tracking metadata used are those archived in the CCSDS–RDEF files and mission tracking records \citep{Aggarwal2025b, Jain_2024}.

\noindent\textbf{Basic pre-processing:} Raw RDEF binaries were converted to time–tagged carrier streams using mission tracking metadata. Carrier extraction and generation of 1 s spectrogram frames followed the pipeline described in \citet{Aggarwal2025a}. 

\noindent\textbf{Spectral estimation and uncertainty:} Each 1 s spectrogram was modeled with a Gaussian power spectral shape to extract zeroth, first and second moments (total power $P$, Doppler shift $\Delta f$, and spectral width $B_S$) as in \citet{Aggarwal2025a}. A one–minute moving average stabilizes the estimates because solar offset varies slowly over that interval \citep{Woo1979, Coles1991}. Oscillator instability contributes a negligible frequency error (e.g. $\sim10^{-12}\times f \approx 8.4\times10^{-3}$ Hz for X–band); typical statistical scatter in $B_S$ is 15-30\% and dominates the error budget. We combine oscillator and statistical errors in quadrature to produce realistic uncertainties on spectral width and on the propagated velocity estimates.

\noindent\textbf{Science context of the two data sets:} The MOM S–band observations probe the inner heliosphere at $5$-$8 R_{\odot}$ during a relatively quiet phase and provide constraints on low–speed solar wind and coronal electron densities appropriate to near–ecliptic streamers \citep{Jain2022}. The Akatsuki X–band data sample closer to the Sun ($\sim1.4$-$10 R_{\odot}$) and cover both slow and fast wind regimes across different solar cycle phases, enabling comparison of streamer and coronal–hole plasma properties \citep{Chiba2022, Jain_2024}.

\begin{figure*}[!h]
\centering
 \includegraphics[width = 1\linewidth]{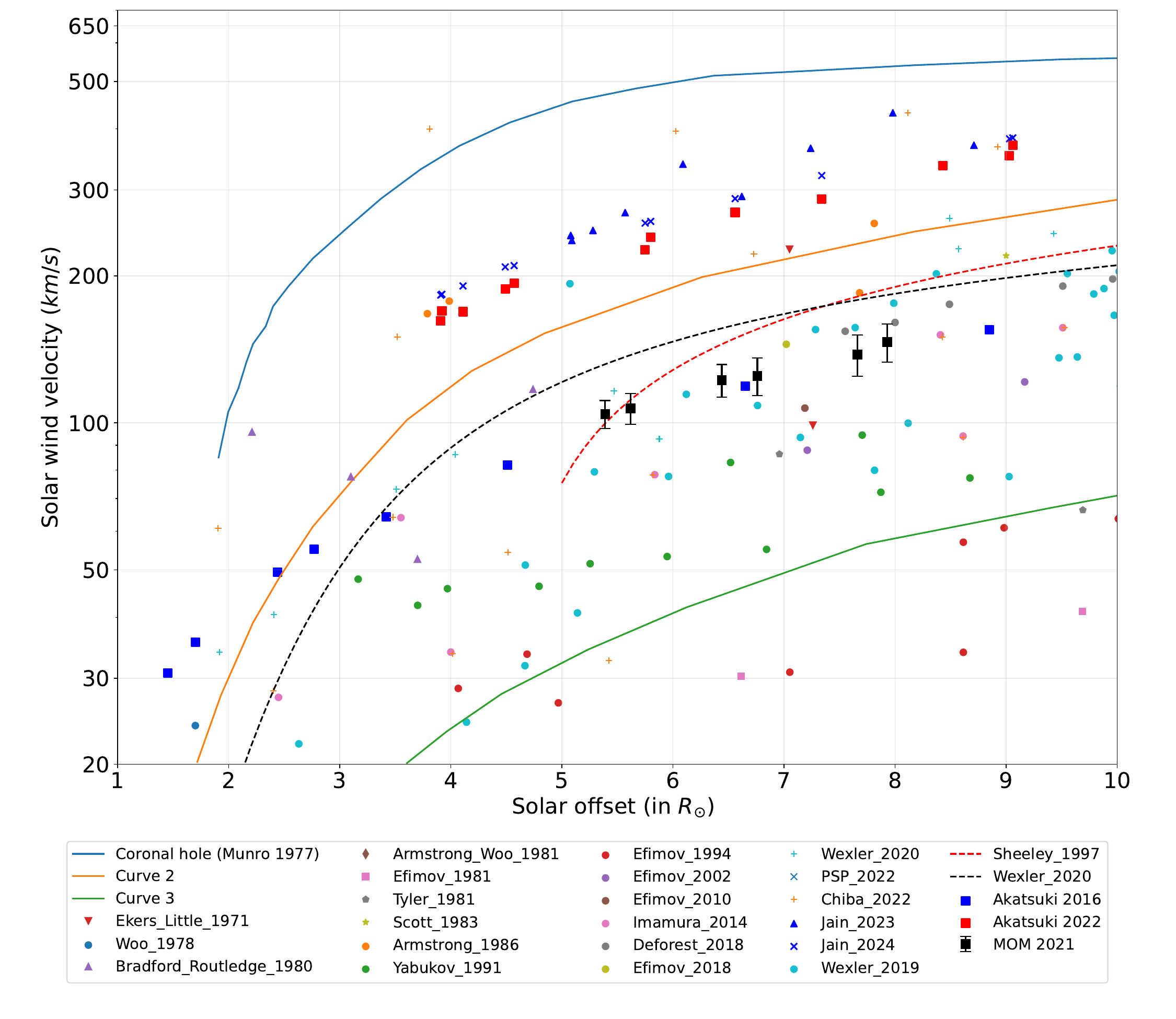}
 \caption{Solar wind speed estimates using the above set of equations for both MOM and Akatsuki RO experiments compared against measurements and models in the literature spanning the past half century.}
 \label{fig:vperp}
\end{figure*}

\subsection{Solar wind and spectral broadening}
\label{sec:methods}
In \citet{Aggarwal2025a}, an empirical formula was found that connects the Doppler spectral width of a radio signal to the solar wind speed perpendicular to the line of sight (LOS). That work was validated for S-band signals. In this paper, we use a simplified version of that relation, which directly links spectral width to solar wind velocity, as in Eq.\,\eqref{method}. Derived wind speeds from past data lay in the range \(100\)–\(150\ \mathrm{km\,s}^{-1}\). The electron densities estimated (on the order of \(10^{10}\ \mathrm{m}^{-3}\)) were consistent with the general form $N_e = N_0 \times \left( \frac{A}{r^{\alpha}} + \frac{B}{r^{\beta}} + \frac{C}{r^{\gamma}} \right)$ shown in earlier models \citep{Edenhofer1977, Esposito1980, Muhleman1981, Strachan1993, Guhathakurta1994, Cranmer1999, Wexler2019}.

Following the approach in \cite{Aggarwal2025a}, one can write the electron density and radial velocity of the solar wind as :
\begin{equation}
    N_e = \frac{f}{r^2 \times R_{SP} \times r_{\odot} \times (1+ R_{SP})} \times \left( \frac{B_s}{c_0} \right)^{5/6}
\end{equation}
\begin{equation}
    v = 1.687 \times \left[ \frac{r\times  R_{EP} \times (1+R_{SP})^2}{R_{SP}} \right] B_s^{\frac{1}{6}}
    \label{method}
\end{equation}

where:
\begin{description}
  \item[\(N_e\)] Electron density in \(\mathrm{cm}^{-3}\)
  \item[\(v\)] radial solar wind velocity in $km s^{-1}$
  \item[\(f\)] Signal frequency in GHz
  \item[\(r\)] Heliocentric distance to the point of closest approach (in \(r_{\odot}\))
  \item[\(R_{SP}\), \(R_{EP}\)] Distance from Sun and Earth to the Probe  (in AU)
  \item[\(B_s\)] Spectral broadening in S-band signal (in Hz)
  \item[\(r_{\odot}\)] Solar radius (in meters)
  \item[\(c_0 = 1.14 \times 10^{-24}\)] Empirical constant (dimensionless)
\end{description}

This formulation assumes a Kolmogorov-type turbulence spectrum (p=11/3) in the coronal plasma and a steady outward solar wind, characteristic of fully developed isotropic turbulence. In this regime, energy cascades from large to small spatial scales without dissipation, leading to a well-defined power-law behavior that governs the observed radio phase and frequency fluctuations. The steady solar wind assumption further allows application of the frozen-flow approximation, linking spatial wavenumbers to temporal frequencies through the constant bulk flow velocity of the plasma. In the radial range we examine (1.4–10 \(R_\odot\)), it is generally assumed that the solar wind is nearly radial while the coronal magnetic field is approximately radial \citep{Woo1977, Waldmeier1977}. Under those conditions, the radial velocity (\(v\)) becomes a good proxy for the total bulk speed. The main advantage of this method is that it is straightforward to apply to Doppler spectral data from radio occultation. It relies on quantities we can observe (e.g. the spectral broadening \(B\)), involves  the use of a few free parameters, and its results have been found to agree with independent estimates based on phase scintillation or frequency residuals \citep{Aggarwal2025a, Aggarwal2025b, Jain_2024}.

However, since this method was originally derived for S-band (2.29 GHz), applying it to X-band (8.4 GHz) requires adjustments. At higher frequencies, scattering effects are weaker, and Doppler widening is smaller. Thus the empirical constants need adjustments when moving from S-band to X-band. We discuss those modifications and validate them using data from Akatsuki in later sections. The theoretical justification and validation of these formulas are also presented in \citet{Aggarwal2025a, Aggarwal2025b}, the process of which can be summarised as :
\begin{enumerate}
  \item From the received signal, determine the spectral broadening \(B\) and find the point of closest approach to the Sun.  
  \item Use the NASA SPICE toolkit to calculate the relevant geometric distances.  
  \item Estimate \(N_e\) with the empirical relation, assuming a power-law density fluctuation spectrum with index \(p = 11/3\) (Kolmogorov turbulence).  
  \item Use \(B\) along with geometry to compute the solar wind speed \(v\).  
\end{enumerate}

\subsection{Derivation of solar wind speed and density using Spectral analysis : a unified approach}

For the method to be applicable for the other frequencies typically used for RO experiments, we have to scale our method for it to be able to account for the change in the extent in the broadening in the received signal and its dependence on the wavelength of the signal. For this, we have from \cite{Morabito2009},
\begin{equation}
 B \sim\lambda^{\frac{5}{6}}
\end{equation}
where $B$ is the spectral broadening in Hz, and $\lambda$ is the wavelength in cm. This equation can be then re-written as:

\begin{equation}
 \frac{B_s}{B_r} =\left( \frac{\lambda_s}{\lambda_r} \right)^{\frac{5}{6}}
\end{equation}

\begin{equation}
 B_s = B_r \times \left( \frac{\lambda_s}{\lambda_r} \right)^{\frac{5}{6}}
 \label{eq:factor}
\end{equation}

where $\lambda_s$ is the wavelength in cm for the S-band frequency (2.29 GHz) used in the occultation experiment conducted using MOM, and can be taken to be $\lambda_s = 13.2 cm$, treated as a constant. $\lambda_r$ is the wavelength in cm for the signal which we are using for a particular occultation experiment. Additionally, substituting the value of $B_s$ from eq. \ref{eq:factor} in the above set of equations, we have

\begin{equation}
    N_e = \frac{f}{r^2 \times R_{SP} \times r_{\odot} \times (1+ R_{SP})} \times \left( \frac{B_r \times \left( \frac{13.2}{\lambda_r} \right)^{\frac{5}{6}}}{c_0} \right)^{5/6}
\end{equation}
and
\begin{equation}
    v = 1.687 \times \left[ \frac{r\times  R_{EP} \times (1+R_{SP})^2}{R_{SP}} \right] \left[ B_r \times \left( \frac{13.2}{\lambda_r} \right)^{\frac{5}{6}} \right]^{\frac{1}{6}}
\end{equation}

Solving the above equations, assuming a Kolmogorov turbulence spectrum with $p = 11/3$ we have the relationships between broadening in the received signal and the electron density $N_e$, and radial wind speeds $v$ as:

\begin{equation}
    N_e = \frac{13.2 \times f}{r^2 \times R_{SP} \times r_{\odot} \times (1+ R_{SP})} \times \left( \frac{B_r}{c_0} \right)^{5/6} \times \frac{1}{\lambda_r}
    \label{eq:final_den}
\end{equation}
and
\begin{equation}
    v = 2.826 \times \left[ \frac{r\times  R_{EP} \times (1+R_{SP})^2}{R_{SP}} \right] B_r^{\frac{1}{6}} \times \left( \frac{1}{\lambda_r} \right)^{\frac{1}{5}}
    \label{eq:final_vel}
\end{equation}

Equations \ref{eq:final_den} and \ref{eq:final_vel} can be used to estimate the Solar wind density and velocity in the near-Sun environment using the radio occultation technique for all frequencies typically used for these experiments to a reasonable extent, assuming Kolmogorov turbulence. 

\begin{table*}
    \centering
    \setlength{\tabcolsep}{3pt}
    \begin{tabular}{|l|l|l|l|l|l|l|l|l|}
    \hline
        Date of& Mission & Freq. & $r$ & $B_{S/X}$ & $V$ & Error & $N_e$ & Error in \\ 
        experiment& Name & (S/X) & ($R_{\odot}$) & Hz & (km s$^{-1}$) & (\%) & (m$^{-3}$) & $N_e$ $(m^{-3}$)\\ \hline
        02/10/21 & MOM & S & 7.93 & 1.15 & 146.35 & 8.96 & $1.08\times10^{10}$ & $3.11\times10^{9}$ \\ 
        03/10/21 & MOM & S & 6.76 & 1.04 & 124.83 & 8.78 & $9.21\times10^{9}$ & $3.68\times10^{9}$ \\ 
        04/10/21 & MOM & S & 5.62 & 1.07 & 107.07 & 7.27 & $1.36\times10^{10}$ & $4.38\times10^{9}$ \\ 
        12/10/21 & MOM & S & 5.39 & 1.06 & 104.27 & 6.57 & $1.27\times10^{10}$ & $4.67\times10^{9}$ \\ 
        13/10/21 & MOM & S & 6.49 & 1.24 & 122.37 & 7.64 & $1.46\times10^{10}$ & $3.93\times10^{9}$ \\ 
        14/10/21 & MOM & S & 7.66 & 1.19 & 138.07 & 9.75 & $1.09\times10^{10}$ & $3.36\times10^{9}$ \\ 
        \hline
        04/06/16 & VCO & X & 2.77 & 3.54 & 55.08 & 0.62 & $6.01\times10^{11}$ & $2.23\times10^{10}$ \\ 
        05/06/16 & VCO & X & 1.70 & 4.07 & 35.58 & 0.51 & $1.12\times10^{12}$ & $3.43\times10^{10}$ \\ 
        08/06/16 & VCO & X & 1.45 & 4.07 & 30.72 & 0.49 & $1.31\times10^{12}$ & $4.01\times10^{10}$ \\ 
        09/06/16 & VCO & X & 2.44 & 3.98 & 49.46 & 0.55 & $7.66\times10^{11}$ & $2.52\times10^{10}$ \\ 
        10/06/16 & VCO & X & 3.42 & 2.60 & 64.28 & 1.03 & $3.56\times10^{11}$ & $2.27\times10^{10}$ \\ 
        11/06/16 & VCO & X & 4.51 & 2.25 & 81.90 & 1.20 & $2.35\times10^{11}$ & $1.73\times10^{10}$ \\ 
        13/06/16 & VCO & X & 6.65 & 2.25 & 118.96 & 0.92 & $1.60\times10^{11}$ & $9.05\times10^{9}$ \\ 
        15/06/16 & VCO & X & 8.85 & 1.99 & 155.07 & 1.04 & $1.07\times10^{11}$ & $6.85\times10^{9}$ \\ 
            \hline
        15/10/22 & VCO & X & 9.03 & 1.78 & 352.36 & 0.56 & $3.47\times10^{10}$ & $8.86\times10^{8}$ \\ 
        17/10/22 & VCO & X & 7.34 & 1.92 & 287.43 & 0.96 & $4.58\times10^{10}$ & $1.69\times10^{9}$ \\ 
        19/10/22 & VCO & X & 5.75 & 1.96 & 226.17 & 1.02 & $5.94\times10^{10}$ & $2.28\times10^{9}$ \\ 
        21/10/22 & VCO & X & 4.49 & 2.77 & 188.09 & 1.54 & $1.08\times10^{11}$ & $5.55\times10^{9}$ \\ 
        22/10/22 & VCO & X & 4.11 & 2.43 & 168.93 & 1.56 & $1.03\times10^{11}$ & $5.43\times10^{9}$ \\ 
        23/10/22 & VCO & X & 3.91 & 2.45 & 161.75 & 1.50 & $1.09\times10^{11}$ & $5.56\times10^{9}$ \\ 
        24/10/22 & VCO & X & 3.92 & 2.94 & 169.42 & 1.75 & $1.31\times10^{11}$ & $7.49\times10^{9}$ \\ 
        26/10/22 & VCO & X & 4.57 & 2.30 & 193.05 & 1.60 & $8.76\times10^{10}$ & $4.77\times10^{9}$ \\ 
        28/10/22 & VCO & X & 5.80 & 1.98 & 240.16 & 1.28 & $5.96\times10^{10}$ & $2.75\times10^{9}$ \\ 
        29/10/22 & VCO & X & 6.56 & 1.89 & 269.75 & 0.92 & $5.05\times10^{10}$ & $1.81\times10^{9}$ \\ 
        31/10/22 & VCO & X & 8.43 & 1.85 & 336.08 & 0.88 & $3.86\times10^{10}$ & $1.35\times10^{9}$ \\ 
        01/11/22 & VCO & X & 9.06 & 1.95 & 370.38 & 0.95 & $3.80\times10^{10}$ & $1.39\times10^{9}$ \\ \hline
    \end{tabular}
    \caption{Summary of derived solar wind velocities and plasma parameters from radio occultation experiments. MOM = Mars Orbiter Mission; VCO = Akatsuki (Venus Climate Orbiter). $B_{S/X}$ = spectral broadening; $N_e$ = electron density; $N_{e,\mathrm{err}}$ = uncertainty. Velocities are obtained from spectral broadening analysis; uncertainties are quoted as percentage errors on $V$.}
    \label{tab:summary_extended}
\end{table*}

\section{Results and discussion}

\begin{figure*}[!h]
    \centering
    \includegraphics[width=.75\linewidth]{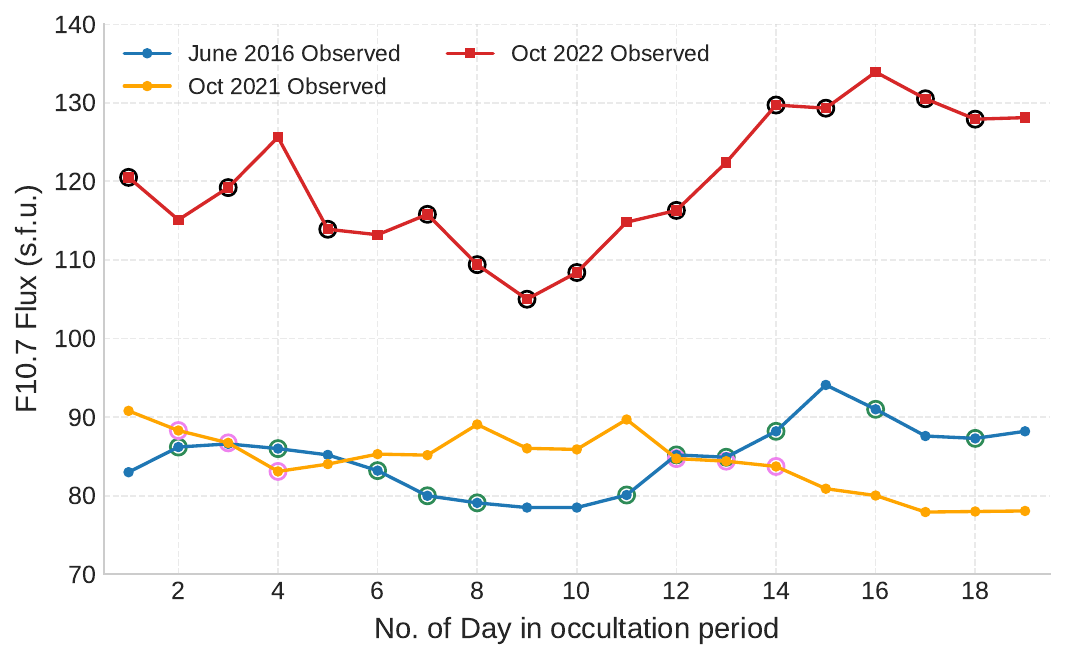}
    \caption{The daily F10.7 values during the period when occultation experiments were conducted in the years 2016, 2021, and 2022 are shown in blue, orange, and red, respectively. The hollow circles on the curves indicate the days on which RO experiments were conducted.}
    \label{fig:flux}
\end{figure*}

The derived spectral broadening parameters from MOM ( 2.29 GHz S-band, October 2021) and Akatsuki (8.41 GHz X-band, June 2016 and October 2022) are summarized in Table \ref{tab:summary_extended}. The corresponding radial dependencies of electron density, $N_e(r)$, and radial solar wind speed, $v(r)$, during these three occultation campaigns are shown in Figure \ref{fig:ne-vperp}. The 2016 Akatsuki observations were conducted during the declining phase of Solar Cycle 24, when solar activity was relatively weak, with an average F10.7 flux of about 80 with solar minimum levels. In contrast, the MOM (2021) and Akatsuki (2022) campaigns occurred during the ascending phase of Solar Cycle 25, characterized by enhanced solar activity and transients, with observation window mean F10.7 flux values of $\sim85$ in 2021 and $\sim120$ in 2022. In figure \ref{fig:flux}, the daily F10.7 values during the period when occultation experiments were conducted in the years 2016, 2021, and 2022 are shown in blue, orange, and red, respectively. The hollow circles on the curves indicate the days on which RO experiments were conducted. Additionally, there were no extreme weather events during the observation period for any of the occultation experiments.

Under the Kolmogorov turbulence framework, the inferred electron densities span $N_e \sim 10^{9}$-$10^{10}$ m$^{-3}$ at heliocentric distances of $r \sim 4$-9 $R_{\odot}$ (Figure \ref{fig:ne-vperp}). Systematic variations are evident across missions, reflecting prevailing solar conditions: the MOM campaign in 2021, under quieter Sun conditions, shows consistently lower $N_e$ and $v$ compared to the more active Akatsuki epochs. Importantly, applying the frequency-scaling formulation causes the S- and X-band results to converge onto a common trend, enabling direct cross-band comparisons.

Figures \ref{fig:ne} and \ref{fig:vperp} place these results in the broader context of earlier radio occultation studies and empirical models. Figure \ref{fig:ne} compares derived coronal densities with observations by \citet{Woo1978} and \citet{Ho2002}, as well as with empirical models by \citet{Baumbach1937, Newkirk1961, Esposito1980, Muhleman1981, Strachan1993, Guhathakurta1994, Ichimoto1996, Leblanc1998, Cranmer1999}. Figure \ref{fig:vperp} provides a comprehensive comparison of our velocity estimates with both observations and models, including those of \citet{Ekers1971, Munro1977, Bradford1980, Armstrong1981, Efimov1981, Tyler1981, Scott1983, Bourgois1985, Armstrong1986, Yakubov1991, Efimov1994, Sheeley1997, Efimov2002, Efimov2010, Imamura2014, DeForest2018, Efimov2018, Wexler2019, Wexler2020, Wexler_2020, Chiba2022, Badman2023, Jain_2024, Aggarwal2025a, Aggarwal2025b}. Additional models for the velocity of the solar wind with and without contributions from MHD waves are represented by curves 2 and 3 \citet{Esser1986}. We observe a decrease in the density with solar offset, which has been seen in previous studies. However, it is to be noted that the density curve is flatter than expected in the lower offsets due to the assumption of $p=11/3$ throughout the observation range, whereas the turbulence slope is flatter in this region \citep{Wexler2020}.

The error budget includes contributions from instrumental noise, oscillator stability, finite integration time, line-of-sight geometry, and turbulence model assumptions. Instrumental instabilities are negligible compared to statistical scatter, while the dominant systematic arises from the assumed Kolmogorov spectrum, which can shift densities and velocities by 10-20\%. The uncertainties reported in Table \ref{tab:summary_extended} capture the main contributions and are representative of the measurement precision achieved.

The main limitations of the approach arise from its underlying assumptions and observational constraints. The method is derived under the assumption of a Kolmogorov inertial range and spherically symmetric density fluctuations. It is worth mentioning that deviations from $p = 11/3$ would modify the proportionality constants in the inferred speed and density relations. The current study only provides an internally consistent framework under these assumptions and relies heavily on the work by \cite{Ho2002} defining empirical relationships using the method of data fitting which uses the work by \cite{Woo1977, Woo1978, Woo1977b}. Regions dominated by anisotropic, evolving, or non-inertial turbulence may require a generalized spectral treatment. 

While the Kolmogorov turbulence model with a fixed $p=11/3$ is used, and it may not fully capture the spectral complexity of coronal electron density fluctuations; especially in regions dominated by anisotropic or non-inertial turbulence; this highlights the importance of simultaneous multi-frequency radio occultation observations. Such measurements would enable more robust validation of frequency-scaling relations and reduce uncertainties in the parameters derived.  Future work would involve working on estimating the local spectral index from the observed frequency-fluctuation power spectra and incorporate this into the empirical relationships, which is expected to significantly reduce these deviations and improve the accuracy of the derived plasma parameters. Such measurements would also improve the physical realism of the framework by accounting for possible departures from the Kolmogorov spectrum and other coronal effects.

An additional point of improvement to be worked on further is the deviation from mass flux conservation in our results which arises from a combination of physical and methodological factors. The apparent deviations from strict mass flux conservation in our derived quantities likely reflect departures from spherical symmetry, localized acceleration effects, and measurement uncertainties, rather than a breakdown of the mathematical formulation itself. Importantly, the simplified model is expected to be most reliable in regions where the solar wind is approximately steady and radially expanding, and less accurate in strongly accelerating or structurally complex regions. 

Finally, from a physical standpoint, within the inner corona below the Alfv\'en critical surface, the solar wind is still undergoing significant acceleration, and the radial electron density does not decrease simply as $1/r^2$ \citep{Leer1980, Stansby2021}. In this acceleration regime, continued energy and momentum deposition modifies both the velocity and density profiles, such that the classical steady-state mass conservation relation is not expected to hold exactly. As a result, the observed non-constancy of $ r^2N_eV$ between $\sim 4$ and $8R_\odot$ reflects a combination of evolving turbulence characteristics, methodological assumptions, and physical deviation from asymptotic solar wind behavior.

Overall, the derived profiles show slopes and magnitudes consistent with classical empirical density models \citep{Woo1978, Edenhofer1977, Esposito1980, Muhleman1981, Strachan1993, Guhathakurta1994, Leblanc1998, Ichimoto1996, Newkirk1961, Baumbach1937, Cranmer1999, Ho2002, Wexler2019} and are in qualitative agreement with in-situ trends from Parker Solar Probe and Solar Orbiter \citep{Badman2023, Marirrodriga2021}. This consistency supports the conclusion that radio occultation, when analyzed with generalized frequency-scaling, provides a reliable remote diagnostic of both coronal density and solar wind acceleration.

\section{Conclusions}
In this study, building on our earlier work, we have developed a versatile, frequency-independent formulation for estimating solar wind velocities and coronal electron densities from radio occultation data across different telecommunication bands of frequencies. The model is developed under the assumptions of (i) a spherically symmetric coronal density distribution, (ii) quasi-steady radial outflow, and (iii) a Kolmogorov inertial-range turbulence spectrum with fixed spectral index $p = 11/3$, and is expected to be most applicable beyond the primary acceleration region and in intervals where turbulence approximates an inertial-range cascade. Its use in regions of strong acceleration or non-Kolmogorov turbulence should be interpreted cautiously. Results from MOM (S-band) and Akatsuki (X-band) show that, once frequency scaling based on the Kolmogorov turbulence assumption is applied, the derived radial trends of $N_e(r)$ and $v(r)$ are mutually consistent and in agreement with empirical coronal models as well as in-situ measurements from inner-heliospheric spacecraft. This demonstrates the method's utility for remote probing of the near-Sun corona and for investigating solar wind acceleration under varying coronal conditions. 

These advances would provide stronger tests of plasma turbulence theories, allow more precise characterization of solar wind acceleration, and deepen our understanding of coronal dynamics. Overall, the unified framework presented here establishes a foundation for flexible and widely applicable remote sensing of the inner heliosphere, paving the way for improved diagnostics of coronal plasma properties in upcoming planetary and solar missions.

\section*{Acknowledgements}
We would like to thank the MOM and Akatsuki mission teams for monitoring the radio signals. K.A. sincerely thanks Director, SPL and Head, HRDD, VSSC, ISRO for the opportunity to visit SPL as part of this project. The Prime Minister's Research scholarship (PMRF) program, Ministry of Education, Government of India, awarded author KA a research scholarship (PMRF-2103356). The authors KA and AD acknowledge the use of facilities procured through the funding via the Department of Science and Technology, Government of India sponsored DST-FIST grant no. SR/FST/PSII/2021/162 (C) awarded to the DAASE, IIT Indore.

\bibliographystyle{elsarticle-harv} 
\bibliography{paper3b}

\end{document}